\documentclass[journal]{IEEEtran}

\ifCLASSINFOpdf
\else
\fi
\usepackage{graphicx} 
\usepackage{float} 
\usepackage{subfigure} 
\usepackage{diagbox}
\usepackage{algorithm}
\usepackage{algorithmic}
\usepackage{multirow}
\usepackage{makecell}
\usepackage{amsmath}
\usepackage{caption}
\usepackage{cite}
\usepackage{color}
\usepackage{url,array}
\usepackage{amssymb}
\usepackage{hyperref}
\usepackage{booktabs}
\usepackage{amsmath}
\usepackage{hyperref}
\usepackage{bbding}
\usepackage{algorithm}
\usepackage{algorithmic}
\newcommand{\PreserveBackslash}[1]{\let\temp=\\#1\let\\=\temp}
\newcolumntype{C}[1]{>{\PreserveBackslash\centering}p{#1}}
\newcolumntype{R}[1]{>{\PreserveBackslash\raggedleft}p{#1}}
\newcolumntype{L}[1]{>{\PreserveBackslash\raggedright}p{#1}}
\begin{document}
%
\title{MUSA: Multi-lingual Speaker Anonymization via Serial Disentanglement}
%
%
%

\author{Jixun~Yao, Qing Wang, Pengcheng Guo, Ziqian Ning, Yuguang Yang, Yu Pan, and~Lei~Xie,~\IEEEmembership{Senior member,~IEEE,}}

%
%

\markboth{Journal of \LaTeX\ Class Files,~Vol.~14, No.~8, August~2015}%
{Shell \MakeLowercase{\textit{et al.}}: Bare Demo of IEEEtran.cls for IEEE Journals}
%



\maketitle

\begin{abstract}
Speaker anonymization is an effective privacy protection solution designed to conceal the speaker's identity while preserving the linguistic content and para-linguistic information of the original speech. While most prior studies focus solely on a single language, an ideal speaker anonymization system should be capable of handling multiple languages. This paper proposes MUSA, a \textbf{MU}lti-lingual \textbf{S}peaker \textbf{A}nonymization approach that employs a serial disentanglement strategy to perform a step-by-step disentanglement from a global time-invariant representation to a temporal time-variant representation. By utilizing semantic distillation and self-supervised speaker distillation, the serial disentanglement strategy can avoid strong inductive biases and exhibit superior generalization performance across different languages. Meanwhile, we propose a straightforward anonymization strategy that employs empty embedding with zero values to simulate the speaker identity concealment process, eliminating the need for conversion to a pseudo-speaker identity and thereby reducing the complexity of speaker anonymization process. Experimental results on VoicePrivacy official datasets and multi-lingual datasets demonstrate that MUSA can effectively protect speaker privacy while preserving linguistic content and para-linguistic information.
\end{abstract}


%
\IEEEpeerreviewmaketitle

\section{Introduction}
Speech is the most innate and direct medium of human communication, utilizing language, intonation, and rhythm to enhance communication with depth and vividness.
In recent years, the proliferation of speech data on the internet has grown exponentially, mainly due to the rise of social media. Applications like telecommunication, voice payment, and speech assistants usually store speech data on centralized servers, leaving sensitive data vulnerable to theft by malicious attackers. This data is rich in personal and sensitive information, such as age, health status, and religious beliefs, which can be identified by speech attribute recognition systems \cite{attacker2020,zhou2021resnext,dibazar2002feature}. Consequently, the protection of sensitive speaker identity information within speech data has become a crucial topic in the field of speech processing.

\begin{figure}[ht]
  \centering
  \includegraphics[width=8.5cm]{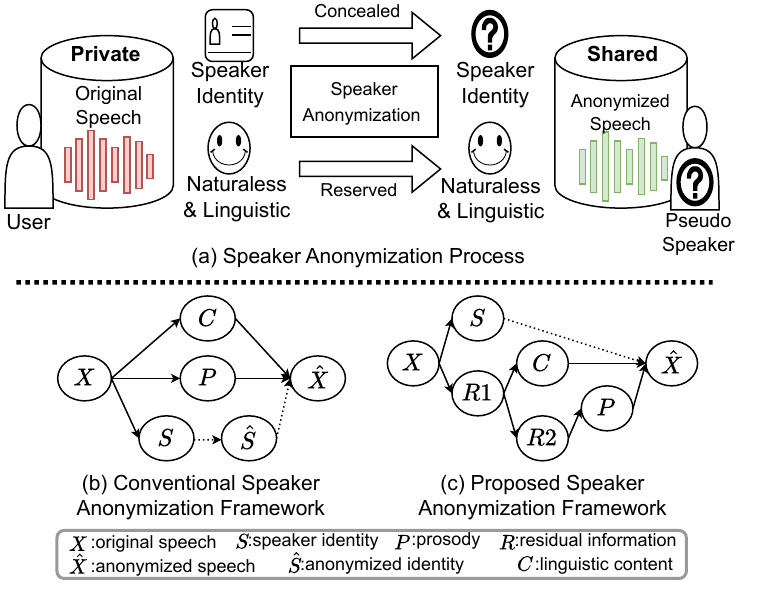}
  \caption{Subfigure (a) illustrates the speaker anonymization process. Subfigure (b) and (c) show the conventional anonymization framework and the proposed framework, respectively. The dashed lines in Subfiguire (b) and (c) represent the anonymization process.}
  \label{fig:vpc}
\end{figure}

In response to recent regulations like the General Data Protection Regulation (GDPR) \cite{nautsch2019gdpr} in the European Union, a new approach to privacy protection, known as \textit{speaker anonymization}, has been proposed. This approach aims to effectively conceal a speaker's identity while maintaining the linguistic content and para-linguistic information (such as prosody, distinctiveness, and naturalness) of the original speech, as illustrated in Figure \ref{fig:vpc}(a). The VoicePrivacy Challenge (VPC) series introduced by the speech community and held in both 2020 \cite{vpc2020eval, vpc2020intro} and 2022 \cite{vpc2022eval}, aims to standardize speaker anonymization tasks and encourage fair comparisons.

Existing speaker anonymization approaches can be categorized into two categories based on the VPC: signal-processing-based anonymization approach \cite{sig1, sig2, sig3,sig4,sig5}, and neural voice conversion-based anonymization approach \cite{fang2019speaker, mawalim2022_speechcom, nn1,nn2,nn3,nn4, nn5, nn10, yao2023distinguishable, f0_anonymization, yao2024taslp, panariello2024speaker}. The neural voice conversion-based anonymization approaches have shown significant advantages over the signal-processing-based ones \cite{vpc2022eval}. Figure \ref{fig:vpc}(b) presents a general framework for a neural voice conversion-based anonymization approach. This approach uses a parallel disentanglement strategy to disentangle the original speech into linguistic content, speaker identity representation, and prosody. It then converts the speaker identity into a pseudo-speaker representation and reconstructs the anonymized speech.
The pseudo-speaker representation is generated by averaging or modifying a set of candidate speaker vectors selected from a speaker vector pool based on distance metrics, thereby protecting the speaker identity information.

While existing approaches can anonymize the original speech to protect speaker identity to some extent, they typically focus on a single language, like English. An ideal speaker anonymization system should be capable of handling multiple languages, even those not included in the training data. To accomplish this, two significant challenges must be addressed. The first is how to disentangle linguistic content, speaker identity, and prosody in multi-lingual scenarios, and adapt effectively to unseen languages. The second challenge is how to effectively conceal speaker identity representation to protect speech privacy.

For the first challenge, utilizing recognition models to explicitly disentangle linguistic content or speaker identity in a parallel way is dependent on prior knowledge and introduces strong inductive biases. This leads to increased complexity and poor adaptability to unseen languages or languages with minimal transcription. Inspired by SpeechTokenizer \cite{zhang2023speechtokenizer}, we present a \textbf{Mu}lti-lingual \textbf{S}peaker \textbf{A}nonymization approach, called MUSA. We propose a serial disentanglement strategy that performs step-by-step disentangling from a global time-invariant representation (\textit{speaker identity}) to a temporal time-variant representation (\textit{linguistic content} and \textit{prosody}), as depicted in Figure \ref{fig:vpc}(c). By utilizing a semantic teacher and self-supervised speaker distillation, the disentangled representation can circumvent strong inductive biases and demonstrate superior generalization performance.

For the second challenge, speaker anonymization should be viewed as a concealment process, not the previous conversion process. Based on the proposed disentanglement strategy, we directly reconstruct the anonymized speech using only the disentangled linguistic content and prosody, both devoid of speaker-related information. The original speaker embeddings are substituted with empty embeddings containing zero values. This simulates the process of concealing speaker identity while eliminating the need for an additional speaker vector pool and reducing complexity. Our extensive evaluations in the multi-lingual speaker anonymization task show that MUSA can effectively protect speaker identity while maintaining original linguistic content and prosody.
The main contributions of this study can be summarized as follows:

We summarize our main contributions as follows:
\begin{itemize}
\item We present MUSA, a multi-lingual speaker anonymization approach. To the best of the author's knowledge, this is the first speaker anonymization approach capable of anonymizing speech across multiple languages and adapting well to unseen languages.
\item We propose a novel anonymization strategy that directly conceals the original speaker identity representation using empty embeddings with zero values. This eliminates the need for conversion to a pseudo-speaker representation and speaker vector pools, reducing the complexity.
\item We introduce two types of distillation methods to disentangle the global time-invariant representation and temporal time-variant representation, which can circumvent strong inductive bias, thereby improving the generalization of serial disentanglement.

\end{itemize}

The rest of the paper is organized as follows.
In Section \uppercase\expandafter{\romannumeral2}, related works are introduced. 
In Section \uppercase\expandafter{\romannumeral3}, we detail the proposed timbre-reserved adversarial attack in the SID system.
Datasets and experimental setup are described in Section \uppercase\expandafter{\romannumeral4}.
Section \uppercase\expandafter{\romannumeral5} presents the experimental results and analysis.
We conclude in Section \uppercase\expandafter{\romannumeral6}.

\section{Related Work}
\subsection{VoicePrivacy Challenge}
The VPC series \cite{vpc2020eval,vpc2022eval} has provided common datasets, evaluation metrics, and baselines to precisely define the speaker anonymization task and ensure a fair comparison. 
According to the VPC, a speaker anonymization system should: 1) output an anonymized speech waveform; 2) conceal the speaker’s identity from different attackers; 3) keep linguistic content and other paralinguistic attributes, e.g. prosody, unchanged to maintain intelligibility and naturalness; 4) ensure all test trials from the same speaker are attributed to the same pseudo-speaker, while test trials from different speakers have different pseudo-speakers.
At the same time, to assess the robustness of anonymization systems under different attack scenarios, it is assumed that attackers have access to a few original or anonymized utterances for each speaker, known as enrollment utterances. Furthermore, attackers may have varying degrees of knowledge about the anonymization system. 

\begin{itemize}
\item Ignorant: Attackers are unaware of the anonymization system, they utilize the original data as enrollment to infer a speaker's identity. 
\item Lazy-informed: Attackers use a similar speaker anonymization system without accurate parameters to anonymize the original data and use the anonymized speech as enrollment to infer a speaker’s identity.
\item Semi-informed: The difference from Lazy-informed is that attackers fine-tune the speaker recognition system using anonymized speech to reduce the mismatch.
\end{itemize}



\subsection{Neural Voice Conversion-based Anonymization Approaches}
A typical approach \cite{fang2019speaker} to speaker anonymization involves using pre-trained automatic speaker verification (ASV) and automatic speech recognition (ASR) models to extract speaker identity and linguistic content, respectively. The original speaker identity is then replaced by averaging a set of candidate speaker identities from an external pool. The key point of speaker anonymization lies in generating a pseudo-speaker identity for the original speaker. Several works \cite{adv_anon,gan_anon} have utilized adversarial training to generate a pseudo-speaker identity, intending to suppress speaker-related information. Furthermore, \cite{mawalim2022_speechcom} and \cite{pca_anon} model a speaker-level latent space and employ techniques such as Principal Component Analysis (PCA) or Singular Value Decomposition (SVD) to obtain a pseudo-identity. \cite{yao_vpc2022} introduced an alternative approach using a look-up-table to generate anonymized speaker embeddings. However, these approaches primarily focus on speaker anonymization in English and are not equipped to anonymize speech data from other languages. 

\begin{figure*}[ht]
  \centering
  \includegraphics[width=17cm]{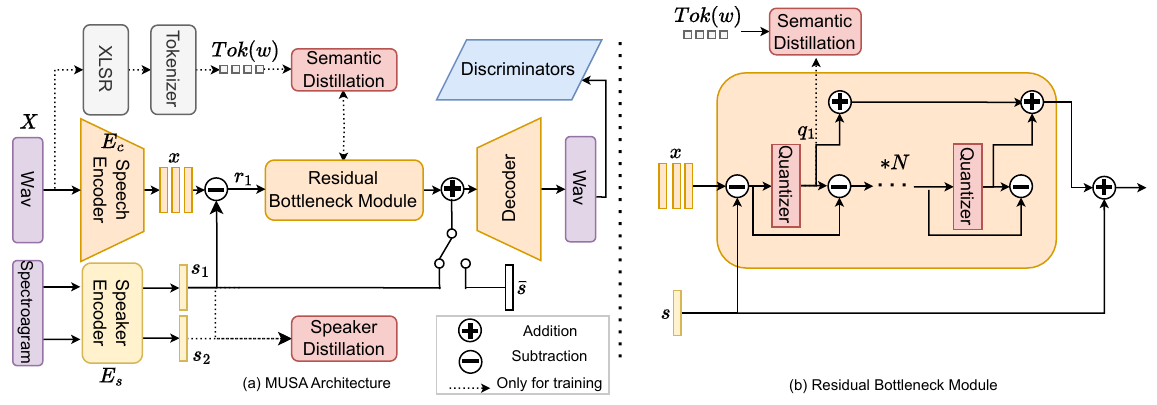}
  \caption{ Subfigure (a) provides an overview architecture of our proposed approach, while subfigure (b) offers details of the residual bottleneck modules. The dashed line is only used in the training process.}

  \label{fig:model}
\end{figure*}

\subsection{Language-independent Speaker Anonymization}
Recent advancements in self-supervised learning (SSL) have demonstrated remarkable performance in speech synthesis. \cite{miao2023anon_conf} uses a soft content representation to replace language-dependent bottleneck features extracted from a pre-trained ASR model, which could anonymize both English and Mandarin speech. Furthermore, \cite{miao2023anon_taslp} proposes an anonymization approach based on an orthogonal Householder neural network to anonymize unseen Mandarin speech data. However, these approaches are only capable of anonymizing two languages and are not equipped to handle more languages. Additionally, the speech content of the anonymized speech can become distorted when the input speech is from an unseen language in the training data. 


\section{Methods}

\subsection{Overview}
As illustrated in Figure\ref{fig:model}(a), MUSA utilizes an auto-encoder architecture, which contains a speech encoder, a speaker encoder, a decoder, and a residual bottleneck module. 

As high-dimensional time series data, speech waveforms can pose computational complexity and resource consumption issues when directly modeled and processed. To address this, we use a speech encoder to capture a frame-level speech representation from the utterance-level speech waveform. These frame-level representations are more directly linked to speech information, including linguistic content, speaker identity, and prosody. 

We then serial disentangle each factor, ultimately reconstructing the original waveform. The serial disentangle process can be described as follows: 1) a speaker encoder is used to extract a time-invariant representation of speaker identity, which is then subtracted from the frame-level representation to disentangle the speaker identity; 2) the representation obtained from subtraction then passes through a residual bottleneck module to disentangle linguistic content and prosody; 3) each disentangled factor is concatenated and used to reconstruct the original speech waveform through the decoder. We will introduce the disentanglement details and two key distillation methods in the below sections. 

\subsection{Speaker Distillation}
In this section, we will introduce the process of speaker distillation. This is a crucial step in extracting the speaker's identity representation and disentangling it from the output of the speech encoder. Let's denote the speech waveform as $X \in \mathbb{R}^T$, where $T$ is the length of the waveform. 
The first step involves feeding $X$ into the speech encoder $E_c$. This allows us to obtain a frame-level speech representation $x \in \mathbb{R}^{d*t}$, where $d$ represents the dimension of the latent representation and $t$ represents the frame length. 

Indeed, the speaker's identity is an inherent and time-invariant characteristic of a voice. We first transform the speech into a spectrogram in our speaker encoder $E_s$. The spectrogram is then fed into $E_s$ and produces a global representation $s \in \mathbb{R}^{d}$ containing time-invariant latent representations after an average pooling. $E_s$ effectively captures the unique identity of the speaker while ensuring that it remains consistent over time.

To ensure the speaker encoder captures more global information and includes details related to speaker identity, we introduce two distillation losses: \textit{segment consistency loss} and \textit{speaker identity loss}. For segment consistency loss, we randomly sample two segments, $\text{seg}_1$ and $\text{seg}_2$, from a same utterance. These segments are processed by $E_s$ to generate two global representations, $s_1$ and $s_2$. Since the speaker's identity is time-invariant, $s_1$ should be similar to $s_2$. We employ cosine similarity as a consistency loss to measure their similarity in a self-supervised manner. The objective is to ensure that these two global representations contain similar information. For speaker identity loss, we use explicit labels $I$ to align the extracted global representation with the speaker's identity. This ensures that the speaker's unique characteristics are accurately captured and represented in the global information. 
The overall speaker distillation loss $\mathcal{L}_{\text{spk}}$ can be defined as follows:
\begin{align}
    s_1, s_2 =& E_s(\text{seg}_1, \text{seg}_2) \\
    \mathcal{L}_{\textrm{spk}}=\mathbb{E}[-log&(C(I\mid s_*))]-\text{cos}(s_1, s_2), 
\end{align}
where $\text{cos}(\cdot)$ denotes cosine similarity and $C(\cdot)$ denotes a speaker identity classifier tasked with determining whether $s_*$ is associated with the corresponding speaker identity.

Thus, we obtain the speaker identity representation $s_1$ or $s_1$ and can further disentangle the speaker identity representation from $x$ to obtain speaker identity independent representation $r_1$ as follows:
\begin{equation}
    r_1 = x - s_1.
\end{equation}
In the next section, we will further disentangle the linguistic content and prosody from $r_1$.

\subsection{Semantic Distillation}
To further disentangle linguistic content and prosody from $r_1$, we utilize a residual bottleneck module for hierarchical disentanglement, as shown in Figure \ref{fig:model}(b). This module consists of $N$ Vector Quantization (VQ) layers, with each VQ layer cascading in a residual manner. The unquantized vector, i.e. $x$, is passed through the first quantizer, and quantization residuals are computed. These residuals are then iteratively quantized by a sequence of $N-1$ additional quantizers. As a result, each subsequent quantizer only contains residual information from the preceding quantizer, making it inherently suitable for disentanglement. This architecture ensures efficient and effective hierarchical disentanglement.


To disentangle diverse information across different VQ layers, we introduce a semantic distillation for the first quantizer. This distillation disentangles the linguistic content from $r_1$, while the supplement information of the other quantizers is assumed to be prosody. We use XLS-R \cite{xlsr_2022}, a large-scale pre-trained model on a multi-lingual dataset~\footnote{\url{https://huggingface.co/facebook/wav2vec2-xls-r-300m}}, as the semantic teacher. Given that the output of the first quantizer is a discrete token, we employ a feature tokenizer to quantize the last layer feature of XLS-R, thereby obtaining the discrete token. The last layer features are closer to the final constraints during training XLS-R, thus containing more linguistic content information.

Conventional feature tokenizers use the K-means clustering approach, which may result in a loss of semantic information compared to the original XLS-R feature. Moreover, not all sets of hidden features are suitable for clustering. Inspired by \cite{huang2023repcodec}, we leverage a neural codec containing a single quantizer to quantize the XLS-R feature. 
This approach helps to preserve more semantic information, enhancing the effectiveness of semantic distillation.

Let $q_i$ represent the outputs of the $i$th quantizer and $\text{Tok}(\cdot)$ denote the feature tokenizer. The semantic distillation loss $\mathcal{L}_{\textrm{sem}}$ can be described as follows:
\begin{equation}
    \mathcal{L}_{\textrm{sem}}=\mathbb{E}[-log(\text{Tok}(w)\mid q_1))],
\end{equation}
where $w$ denotes the last layer feature of the XLS-R model. Semantic distillation forces the first quantizer to effectively encode content information, thus the residual information of the first quantize $r_2$ should contain prosody information. The subsequent quantizers, equipped with a residual structure, supplement the remaining prosody information ($q_2$ to $q_N$). Meanwhile, semantic distillation also can prevent the leakage of speaker identity information to $r_1$.

These two distillation processes sequentially disentangle the speaker's identity, linguistic content, and prosody from the frame-level speech representation. Finally, the speaker identity representation is added to the output of the residual bottleneck module, and the reconstructed waveform is predicted through the decoder.


\subsection{Training and Anonymization Process}
\textbf{Training process}. The training objective of our approach includes a fundamental reconstruction task, a speaker distillation task, and a semantic distillation task. Regarding the reconstruction task, we follow the conventional autoencoder training objective. This objective optimizes a combination of a reconstruction loss, a discriminative loss, and a VQ commitment loss. 

The reconstruction loss consists of both time and frequency domain losses. In the time domain, we aim to minimize the L1 distance between the input waveform $X$ and the reconstructed waveform $\hat{X}$. In the frequency domain, we integrate the L1 and L2 losses over the mel-spectrogram scale. The total reconstruction losses are as follows:
\begin{align}
    \mathcal{L}_{\textrm{rec}}=\Vert X-\hat{X}\Vert_1+\Vert\text{mel}(X)-\text{mel}(\hat{X})\Vert_1 \notag \\ 
    +\Vert\text{mel}(X)-\text{mel}(\hat{X})\Vert_2,
\end{align}
where $\text{mel}(\cdot)$ is an 80-bin mel-spectrogram process using a short-time Fourier transform (STFT).

As for discriminators, we employ the same ones as HiFi-Codec \cite{yang2023hificodec}. These include a multi-scale STFT-based discriminator, a multi-period discriminator, and a multi-scale discriminator. The discriminative loss is denoted as:
\begin{align}
    \mathcal{L}_{\textrm{adv}}(D)=& \mathbb{E}\left[(D(X)-1)^2+D(\hat{X})^2\right] \\
    \mathcal{L}_{\textrm{adv}}(G)=& \mathbb{E}\left[(D(\hat{X})-1)^2\right],
\end{align}
where $G$ and $D$ represent the generator and discriminator, respectively. A feature matching loss is used to measure the difference in the feature of the discriminator between a ground truth sample and a reconstructed sample, which defined as
\begin{equation}
    \mathcal{L}_{\textrm{fm}}(G,D)=\mathbb{E}\left[ \sum_{i=1}^T \frac{1}{N_i}\left\|D^i(X)-D^i(\hat{X})\right\|_1 \right]
\end{equation}
where $D^i$ and $N_i$ denote the features and the number of features in the $i$-th layer of the discriminator, while $T$ denotes the number of layers in the discriminator.

Meanwhile, similar to the conventional VQ models, we employ a straight-through estimator to optimize the commitment loss between the input feature and quantized feature:
\begin{equation}
    \mathcal{L}_{\textrm{com}}=\sum_{i=1}^{N}\left\|x_i-q_i\right\|_2^2,
\end{equation}
Generally, our proposed approach is trained to optimize the mixture of the following losses:
\begin{align}
\mathcal{L}=\lambda_r\mathcal{L}_{\textrm{rec}}
    +\lambda_a\mathcal{L}_{\textrm{adv}}
    +\lambda_f\mathcal{L}_{\textrm{fm}}
    +\lambda_c\mathcal{L}_{\textrm{com}} \notag\\ 
    +\lambda_s\mathcal{L}_{\textrm{spk}}
    +\lambda_m\mathcal{L}_{\textrm{sem}},
\end{align}
where $\lambda_r$, $\lambda_a$, $\lambda_f$, $\lambda_c$, $\lambda_s$ and $\lambda_m$ are hyper-parameters used to balance each loss term.

\begin{table}[ht]
\centering
\caption{Comparing requirements and complexity of previous approach vs. MUSA in anonymization process.} \label{tab:complexity}
\resizebox{1.0\linewidth}{!}{
\begin{tabular}{lcc}
\hline
                                        & Previous         & MUSA                        \\ \hline
Pre-trained model                       & ASR and ASV      & \XSolidBrush                         \\ \hline
\multirow{2}{*}{Addtional pool}         & Store candidate  & \multirow{2}{*}{\XSolidBrush}        \\
                                        & speaker vector   &                             \\ \hline
\multirow{2}{*}{Other speaker identity} & Random select    & \multirow{2}{*}{\XSolidBrush}        \\
                                        & from pool        &                             \\ \hline
\multirow{2}{*}{Complexity}             & Search N         & \multirow{2}{*}{Typical VC} \\
                                        & candidate vector &                             \\ \hline
\end{tabular}
}
\end{table}

\textbf{Anonymization process}. Our proposed anonymization strategy, illustrated in Figure~\ref{fig:model}(a), involves replacing $s$ with an empty embedding $\Bar{s}$ containing zero values during the decoder's reconstruction process. This simulates the concealment of speaker identity, thereby achieving anonymization. As a result, the decoder appears to receive only linguistic content and prosody information for reconstructing the final waveform. Additionally, we can use a linear blend of $s$ and $\Bar{s}$ to achieve varying degrees of anonymization, trading off privacy and utility.

Moreover, our anonymization process is considerably simpler than those in previous anonymization approaches. The comparison results, shown in Table \ref{tab:complexity}, show that the proposed anonymization strategy does not depend on pre-trained recognition systems or an additional speaker vector pool, a crucial component in the previous approach. In contrast, the anonymization process in most of the previous approaches introduces information from other speakers, potentially compromising the voice's uniqueness. Regarding complexity, our approach's anonymization process is a typical VC process, not a search for the $N$ farthest speaker vectors to generate a pseudo-speaker.

\section{Experiments Setup}
\subsection{Datasets}
To ensure a comprehensive and fair comparison, we separately train MUSA using both English and multi-lingual datasets, the trained model can be denoted as $\text{MUSA}_\text{en}$ and $\text{MUSA}_\text{ml}$. The English datasets are constructed from the same VPC standard datasets \cite{vpc2022eval}: LibriTTS-600 (comprising clean-100 and other-500) \cite{libritts} and LibriSpeech-600 (comprising clean-100 and other-500) \cite{librispeech}. For the multi-lingual datasets, we employ Multi-lingual LibriSpeech (MLS) \cite{mls_2020}, a comprehensive multi-lingual dataset encompassing seven languages. We utilize all languages in this dataset, excluding English, details are shown in Figure~\ref{tab:mls_datasets}. We convert the sampling rate of all speech data to 16kHz.

\begin{table}[ht]
\centering
\caption{Details of training and evaluation datasets used in multi-lingual experiments.}\label{tab:mls_datasets}
\resizebox{1.0\linewidth}{!}{
\begin{tabular}{lcccccc}
\hline
\multicolumn{1}{c}{\multirow{2}{*}{Language}} & \multicolumn{3}{c}{Speaker}  & \multicolumn{3}{c}{Duration} \\ \cline{2-7} 
\multicolumn{1}{c}{}                          & train    & dev    & test        & train     & dev    & test    \\ \hline
Dutch                                         & 40       & 6      & 6           & 155       & 12     & 12      \\
French                                        & 142      & 18     & 18          & 107       & 10     & 10      \\
German                                        & 176      & 30     & 30          & 196       & 14     & 14      \\
Italian                                       & 65       & 10     & 10          & 147       & 5      & 5       \\
Polish                                        & 11       & 4      & 4            & 103       & 2      & 2       \\
Portuguese                                    & 42       & 10     & 10           & 160       & 3      & 3       \\
Spanish                                       & 86       & 20     & 20          & 117       & 10     & 10      \\ \hline
\end{tabular}
}
\end{table}

For the evaluation dataset in English, we follow the VPC 2022 configuration \cite{vpc2022eval}, employing LibriSpeech-test-clean and VCTK-test \cite{vctk}. In the evaluation dataset for the multilingual scenario, we utilize MLS test datasets to evaluate the performance of our proposed approach. When evaluating the unseen language scenario, we select Chinese Mandarin as the unseen language and employ AISHELL-3 datasets \cite{shi2020aishell3} to evaluate the anonymization performance.

\subsection{Model Configurations}
\textbf{Speech Encoder}:
The speech encoder is structured with four convolution blocks, each integrating a residual unit followed by a down-sampling layer. Within the residual unit, two convolutions with a kernel size of 3 and a skip connection are employed, while each down-sampling layer entails a convolution layer with a kernel size twice the stride. The number of channels doubles during down-sampling, and the strides for the four convolution blocks are set as (2, 4, 5, 8). After the convolution blocks, there is a two-layer LSTM for sequence modeling and a concluding 1D convolution layer with a kernel size of 7 and 512 output channels.

\textbf{Residual Bottleneck Module}:
We employ 8 quantizers in the residual bottleneck module. Each quantizer employs a codebook of size 1024 and is trained using exponential moving average updates, in line with the method proposed in \cite{vq_ref}.

\textbf{Decoder}:
The architecture of the decoder mirrors the encoder, utilizing transposed convolutions in place of stride convolutions, with the strides in the reverse order of those in the encoder. 
Regarding the speaker encoder and discriminators, we follow the same configuration as the StyleSpeech \cite{stylespeech} and HiFi-Codec \cite{yang2023hificodec}, respectively.

\textbf{Model Training}:
Both $\text{MUSA}_\text{en}$ and $\text{MUSA}_\text{ml}$ are trained using the AdamW \cite{adamw} optimizer using parameters $\beta_1$ = 0.8, $\beta_2$ = 0.99, and weight decay $\lambda$ = 0.01. The learning rate decay followed a schedule with a decay factor of 0.999 per epoch, starting from an initial learning rate of $2 \times 10^{-4}$. The training process comprised a total of 500,000 steps, utilizing 8 NVIDIA 3090 GPUs with a batch size of 128 utterances. For the hyper-parameters of the total training loss, we set $\lambda_r=45$, $\lambda_a=1$, $\lambda_f=1$, $\lambda_c=0.1$, $\lambda_s=1$ and $\lambda_m=1$.


\subsection{Evaluation Metrics}
\textbf{Privacy Metric:} To assess the privacy protection capability, we use the equal error rate (EER) as the primary privacy metric, computed by a pre-trained ASV model. We employ the VPC official ASV model and open source pre-trained ECAPA-TDNN~\footnote{\url{https://huggingface.co/speechbrain/spkrec-ecapa-voxceleb}} model \cite{ecapa_tdnn} for English language evaluation and multi-lingual scenarios evaluation, respectively.
EER is defined as the rate at which both false acceptance and false rejection errors are equal:
\begin{equation}
    \text{EER} = P(\text{FPR}=\text{FNR}),
\end{equation}
where FPR (False Positive Rate) is the proportion of negative instances (non-targets) incorrectly classified as positive (targets) and FNR (False Negative Rate) is the proportion of positive instances (targets) incorrectly classified as negative (non-targets).

\textbf{Intelligibility Metric:}
We use word error rate (WER) computed by a pre-trained ASR model for intelligibility evaluation. The VPC official ASV model is employed to evaluate the intelligibility performance in the English language. For multi-lingual and unseen language scenarios, we utilize Whisper-small~\footnote{\url{https://huggingface.co/openai/whisper-small}}\cite{whisper} and WeNet toolkits~\footnote{\url{https://github.com/wenet-e2e}} for intelligibility evaluation.
WER measures the difference between the ASR output and the ground truth transcript:
\begin{equation}
    \text{WER}=\frac{S+D+I}{N},
\end{equation}
where $S$, $D$, and $I$ represent the number of substitutions, deletions, and insertions, respectively. $N$ represents the number of words in the reference answer.

\textbf{Intonation Metric:}
To measure how well anonymization preserves the intonation of the original speech, the pitch correlation metric $\rho^{F0}$ is computed. $\rho^{F0}$ is the Pearson correlation \cite{sedgwick2012pearson} between the pitch sequences of original speech and anonymized speech. 

\textbf{Voice Distinctiveness Metric:}
The gain of voice distinctiveness metric ($G_{\textrm{vd}}$) aims to assess the voice distinctiveness preservation of anonymized speech \cite{gvd1,gvd2}. This metric relies on two voice similarity matrices. A voice similarity matrix $M=(M(i,j))_{1 \leq i \leq N, 1 \leq j \leq N}$ is computed for N speakers and similarity values $M=(M(i,j))$ computed for speakers $i$ and $j$ as follows:

\begin{align}
\resizebox{0.9\linewidth}{!}{$M(i, j)=\operatorname{sigmoid}\left(\frac{1}{n_i n_j} \sum_{\substack{1 \leq k \leq n_i \\ 1 \leq l \leq n_j \\ k \neq l \text { if } i=j}} \operatorname{LLR}\left(x_k^{(i)}, x_l^{(j)}\right)\right)$},
\end{align}
where $\operatorname{LLR}(x_k^{(i)}, x_l^{(j)})$ is the log-likelihood-ratio obtained by comparing the $k$-th segment from the $i$-th speaker with the $l$-th segment from the $j$-th speaker and $n_i$ and $n_j$ are the numbers of utterances for speaker $i$ and speaker $j$, respectively. The log-likelihood ratio is obtained by the ASV model. $M_{oo}$ (for \textbf{o}riginal enrollment) and $M_{aa}$ (for \textbf{a}nonymized enrollment) are prepared for computing the diagonal dominance $D_{\text {diag }}(M)$, which is the absolute difference between the mean values of diagonal and off-diagonal elements, formulated as follows:
\begin{equation}
    D_{\text {diag }}(M)=\left|\sum_{1 \leq i \leq N} \frac{M(i, i)}{N}-\sum_{\substack{1 \leq j \leq N \\ 1 \leq k \leq N \\ j \neq k}} \frac{M(j, k)}{N(N-1)}\right|.
\end{equation}
The gain of voice distinctiveness metric $G_{\textrm{vd}}$ can be mathematically expressed as the ratio between the diagonal dominance of two matrices:
\begin{equation}
    G_{\mathrm{vd}}=10 \log _{10} \frac{D_{\mathrm{diag}}\left(M_{\mathrm{aa}}\right)}{D_{\mathrm{diag}}\left(M_{\mathrm{oo}}\right)},
\end{equation}
when $G_{\mathrm{vd}}=0$ signifies that the voice distinctiveness remains unchanged after the anonymization process. Conversely, a $G_{\mathrm{vd}}$ gain value above or below 0 indicates an average increase or decrease in voice distinctiveness, respectively.

\textbf{Speech Quality Metric:}  To further analyze the speech quality anonymized by our proposed approaches, we utilize a recently proposed mean opinion score (MOS) prediction network, called UTMOS~\cite{utmos}, to estimate the perceived quality, which served as an additional utility metric\footnote{\url{https://github.com/tarepan/SpeechMOS}}.

For a comprehensive evaluation of MUSA, we utilize the VPC official evaluation pipeline \cite{vpc2022eval} for the English language and apply the same metrics for multi-lingual scenarios and unseen language. We assume the anonymization system is a black box for attackers, MUSA is only evaluated in Ignorant and Lazy-informed scenarios. According to the VPC request, the $\rho^{F0}$ of anonymized speech must achieve a minimum average of 0.3.

\subsection{Baseline Systems}
We compare the anonymization performance in the English language with three VPC official baseline systems \cite{vpc2022eval}: \textbf{B1.a}, a two-stage anonymization framework that includes an acoustic model and a vocoder, employing an average anonymization strategy; \textbf{B1.b}, based on the same idea and anonymization strategy as \textbf{B1.a}, with the main difference being the architecture, which is an end-to-end framework; \textbf{B2}, a signal-processing-based approach that employs the McAdams coefficient to achieve anonymization by shifting the pole positions derived from linear predictive coding.

Additionally, we compare the proposed anonymization approach with the first-ranking anonymization system in the VPC 2022, referred to as \textbf{ASV-Free}~\cite{yao_vpc2022}, which uses LUT to replace the x-vector pool and combines two types of speaker embedding to achieve speaker anonymization. 
We also consider an upgraded version of \textbf{ASV-Free} called \textbf{Formant}~\cite{yao2023distinguishable}, which models speaker identity based on formant and F0 rather than LUT, serving as another powerful baseline system.

\begin{table*}[ht]
\centering
\caption{Averaged results over baselines and $\text{MUSA}_\text{en}$ on VPC development and test datasets. The best scores in the column of the test datasets are given in bold. $\text{EER}_{\text{ig}}$ and $\text{EER}_{\text{la}}$ represent the EER results on Ignorant and Lazy-informed scenarios, which closer to 50\% is better. For WER, lower is better. For $\rho^{F0}$, $G_{\textrm{vd}}$ and UTMOS, higher is better. }\label{tab:compare}

\renewcommand\arraystretch{1.3}
\resizebox{0.9\linewidth}{!}{
\begin{tabular}{lcccccccccccc}
\hline
     & \multicolumn{2}{c}{$\text{EER}_{\text{ig}}$ $\uparrow$} & \multicolumn{2}{c}{$\text{EER}_{\text{la}}$ $\uparrow$} & \multicolumn{2}{c}{WER $\downarrow$}  & \multicolumn{2}{c}{$\rho^{F0}$ $\uparrow$} & \multicolumn{2}{c}{$G_{\textrm{vd}}$ $\uparrow$} & \multicolumn{2}{c}{UTMOS $\uparrow$}\\ \cline{2-13} 
     & dev        & test       & dev        & test       & dev        & test       & dev        & test  & dev        & test      & dev        & test       \\ \hline
Orig & 3.54       & 3.79       & 3.54       & 3.79       & 7.30       & 8.48       & -          & -         & 0          & 0   & 4.31       & 4.27       \\
B1.a~\cite{vpc2022eval} & 53.14      & 50.29      & 32.12      & 32.82      & 10.88      & 10.98      & 0.77       & 0.77      & -9.17      & -10.15     & 3.31       & 3.39\\
B1.b~\cite{vpc2022eval} & 53.91      & 52.14      & 27.39      & 27.51      & 10.69      & 10.84      & 0.80       & 0.80      & -6.44      & -6.44      & 3.68       & 3.71\\
B2~\cite{vpc2022eval}   & 37.01      & 38.29      & 43.80      & 45.04      & 17.15      & 18.52      & 0.80       & 0.80      & \textbf{-1.72}      & \textbf{-1.63}      & 3.38       & 3.37\\
ASV-Free~\cite{yao_vpc2022} & 52.11	& 51.27	   & 40.83	    & 39.77	     & \textbf{ 5.63}	       & \textbf{5.86}	   & 0.72	    & 0.70	    & -18.86	 & -18.69      & 3.64       & 3.75\\
Formant~\cite{yao2023distinguishable}  & 49.55	& 48.63	   & 41.22	    & 40.76	     & 6.95	      & 6.82	   & 0.80	    & 0.81	    & -4.55	 & -4.92      & 3.58       & 3.56\\
$\text{MUSA}_\text{en}$ & \textbf{52.49}      & \textbf{54.17}      & \textbf{47.93}      & \textbf{48.65}     & 9.88       & 9.91       & \textbf{0.82}       & \textbf{0.83}      & -2.93      & -2.78      & \textbf{3.83}       & \textbf{3.76}\\ \hline
\end{tabular}}
\end{table*}

\section{Experimental Results}
\subsection{Speaker Anonymization Experiments in English}
We begin by evaluating our approach using English datasets and comparing it with the official VPC baseline systems and the start-of-the-art anonymization systems. This served as the basis for evaluating scenarios involving multiple languages and unseen languages. Table \ref{tab:compare} presents the average results of $\text{MUSA}_\text{en}$ and baseline systems on VPC development and test datasets.

In terms of privacy protection, $\text{MUSA}_\text{en}$ achieves 54.17\% and 48.65\% on test datasets for the Ignorant and Lazy-informed scenarios, respectively. These EER results outperform all five baseline systems. Meanwhile, the EER results of \textbf{B1.a} and \textbf{B1.b} systems experience significant degradation in the Lazy-informed scenario, with a noticeable gap compared to the Ignorant scenario. This indicates that the VPC baseline system is ineffective in protecting privacy when an attacker has advanced knowledge of the anonymization system. In contrast, the EER results of $\text{MUSA}_\text{en}$ only decrease by approximately 5\% in the Lazy-informed scenario, illustrating the effectiveness of our approach in protecting privacy in different attack scenarios.

In the utility evaluation, the WER results of $\text{MUSA}_\text{en}$ are superior to both \textbf{B1.a} and \textbf{B1.b} and significantly lower than \textbf{B2}. In terms of pitch correlation $\rho^{F0}$, $\text{MUSA}_\text{en}$ achieves 0.82 and 0.83 on the development and test datasets, respectively. These pitch correlation scores not only meet the minimum requirement but also outperform all baseline systems. While the $G_{\textrm{vd}}$ results of $\text{MUSA}_\text{en}$ are less than \textbf{B2}, a consideration of the WER results shows that \textbf{B2} is not preferable due to the significant degradation in the intelligibility of anonymized speech. Although \textbf{ASV-Free} achieves the lowest WER results, it falls behind our approach in terms of pitch correlation $\rho^{F0}$ and voice distinctiveness $G_{\textrm{vd}}$, especially in $G_{\textrm{vd}}$. The $G_{\textrm{vd}}$ results indicate that \textbf{ASV-Free} is undesirable, as it produces a similar voice regardless of the original speaker. Meanwhile, the UTMOS results of $\text{MUSA}_\text{en}$ outperform all baseline systems, demonstrating that the speech anonymized by our system has better quality.

\begin{figure*}[ht]
  \centering
  \includegraphics[width=1.0\linewidth]{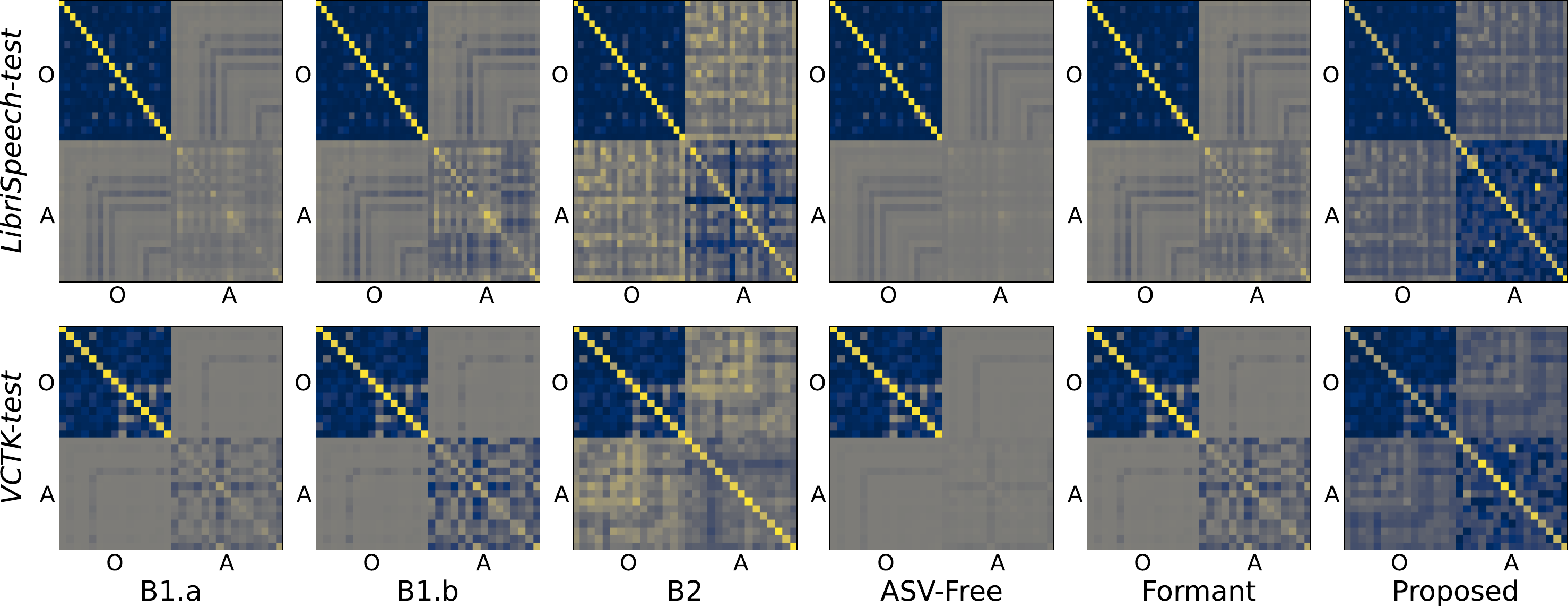}
  \caption{The visualization of the similarity matrix between the original speaker (O) and the anonymized speaker (A). In each matrix, the upper-left submatrix is $M_{oo}$, which is the similarity matrix between the original speaker and the original speaker; the upper-right $M_{oa}$ or lower-left $M_{ao}$ submatrix is the similarity matrix between the original speaker and the anonymized speaker; the lower-right submatrix $M_{aa}$ is the similarity matrix between the anonymized speaker and the anonymized speaker.}
  \label{fig:gvd}

\end{figure*}

To clearly show the comparison of voice distinctiveness performance between $\text{MUSA}_\text{en}$ and the baseline systems, we utilize visualizations of the similarity matrix used to calculate the $G_{\textrm{vd}}$ metrics for the anonymized speech, as depicted in Figure~\ref{fig:gvd}. The similarity matrix is computed using VPC official toolkits on LibriSpeech and VCTK test datasets. The presence of a distinct dominant diagonal in $M_{aa}$ reflects higher voice distinctiveness, which disappears when the speaker is different. In the case of B1.a, B1.b, and Formant, there is an unclear dominant diagonal in $M_{aa}$, indicating a deterioration in voice distinctiveness after the anonymization process. For ASV-Free, the dominant diagonal in $M_{aa}$ disappears, rendering the anonymized speaker indistinguishable and signifying a significant degradation in voice distinctiveness. In contrast, the matrices for B1.b and $\text{MUSA}_\text{en}$ exhibit distinct diagonals in $M_{aa}$, suggesting successful preservation of voice distinctiveness after the anonymization process. The results of the above experiment in terms of privacy and utility illustrate the superior anonymization performance of $\text{MUSA}_\text{en}$ in the English language.

\begin{table}[ht]
\centering
\caption{Averaged results of $\text{MUSA}_\text{ml}$ compared with the original speech in multi-lingual scenarios.}\label{tab:multi-lingual}

\renewcommand\arraystretch{1.3}
\resizebox{1.0\linewidth}{!}{
\begin{tabular}{lccccccc}
\hline
                          &      & $\text{EER}_{\text{ig}}$   & $\text{EER}_{\text{la}}$   & WER   & $\rho^{F0}$   & $G_{\textrm{vd}}$   & UTMOS\\ \hline
\multirow{2}{*}{Orig}     & dev  & 1.83  & 1.83  & 15.81 & -    & 0    & 4.17  \\
                          & test & 2.01  & 2.01  & 15.14 & -    & 0    & 4.13 \\
\multirow{2}{*}{$\text{MUSA}_\text{ml}$} & dev  & 33.82 & 26.59 & 16.96 & 0.74 & -2.35 & 3.72\\
                          & test & 34.88 & 28.76 & 17.06 & 0.73 & -2.73 & 3.77\\ \hline
\end{tabular}
}

\end{table}

\subsection{Speaker Anonymization Experiments in Multi-lingual Scenarios}
Table \ref{tab:multi-lingual} shows the averaged anonymization results of $\text{MUSA}_\text{ml}$ in multi-lingual scenarios compared to the original speech. The first observation is that $\text{MUSA}_\text{ml}$ achieves EERs that are approximately 30\% higher in both Ignorant and Lazy-informed scenarios. This suggests that $\text{MUSA}_\text{ml}$ effectively protects identity privacy across various languages. At the same time, the WER results of the anonymized speech are only 1.92\% higher than those of the original speech in the test datasets and UTMOS results are 3.72 and 3.77, respectively. These results indicate that $\text{MUSA}_\text{ml}$ successfully retains the original linguistic content in multi-lingual scenarios, a challenge for previous parallel-disentangled-based or language-specific-based anonymization approaches. 

Furthermore, in multi-lingual scenarios, $\text{MUSA}_\text{ml}$ achieves $\rho^{F0}$ results of 0.74 and 0.73, which are significantly higher than the minimum correlation requirements of VPC. The $G_{\textrm{vd}}$ results clearly demonstrate that our proposed approach for anonymization speech maintains desirable voice distinctiveness.

\subsection{Speaker Anonymization Experiments in Unseen Language}
On the other hand, we also evaluate the performance of our proposed approach in an unseen language. We test $\text{MUSA}_\text{en}$ and all baseline systems on Mandarin datasets, the results are shown in Table \ref{tab:unseen}. All baseline systems display EERs exceeding 30\% in either Ignorant or Lazy-informed scenarios, but the CERs exceed 40\%. These CER results suggest that the baseline systems achieve a high degree of speaker identity protection by significantly distorting speech content, which is an ineffective strategy for preserving speaker identity privacy. Specifically, the results of baseline imply that employing a language-dependent ASR model trained on the English dataset is unsuitable for extracting linguistic content in an unseen language. This mismatch is the primary cause for the significantly elevated CER results. On the contrary, $\text{MUSA}_\text{en}$ achieves EER results of 29.84\% and 24.68\% in Ignorant and Lazy-informed scenarios, respectively. Although the CERs are 6.71\% higher than the original speech, they are significantly lower than all baseline systems. Moreover, $\text{MUSA}_\text{en}$ outperforms the baseline systems in terms of $\rho^{F0}$ and $G_{\textrm{vd}}$. 

\begin{table}[ht]
\centering
\caption{Averaged results for anonymization in an unseen language of $\text{MUSA}_\text{en}$ and $\text{MUSA}_\text{ml}$ compared to the baseline systems. }\label{tab:unseen}
\renewcommand\arraystretch{1.3}

\resizebox{1.0\linewidth}{!}{
\begin{tabular}{lcccccc}
\hline
     & $\text{EER}_{\text{ig}}$   & $\text{EER}_{\text{la}}$   & CER   & $\rho^{F0}$   & $G_{\textrm{vd}}$   & UTMOS\\ \hline
Orig & 6.32  & 6.32  & 8.97 & -    & 0     & 4.32\\
B1.a & 38.96 & 36.75 & 45.81 & 0.43 & -3.71 & 3.32\\
B1.b & \textbf{40.28} & \textbf{41.51} & 49.77 & 0.38 & -4.06 & 3.43\\
B2   & 34.65 & 32.71 & 40.83 & 0.61 & -3.88 & 3.46\\
ASV-Free   & 39.51 & 38.74 & 38.99 & 0.51 & -12.18 & 3.39\\
Formant   & 40.07 & 39.18 & 41.31 & 0.64 & -6.49 & 3.24\\ \hline
$\text{MUSA}_\text{en}$ & 29.84 & 24.68 & 15.68 & 0.71 & -3.72 & 3.74\\ 
$\text{MUSA}_\text{ml}$ & 30.61 & 24.07 & \textbf{13.39} & \textbf{0.76} & \textbf{-3.65} & \textbf{3.81}\\ \hline
\end{tabular}

}
\end{table}

We further evaluate the anonymization performance of $\text{MUSA}_\text{ml}$ in the unseen language. As shown in Table \ref{tab:unseen}, $\text{MUSA}_\text{ml}$ achieves comparable privacy performance with $\text{MUSA}_\text{en}$ and surpasses it in terms of other metrics. These results demonstrate that our proposed system effectively protects the speaker's identity without significantly compromising the speech content in unseen language scenarios. Furthermore, the CER results indicate that serial disentanglement strategy and semantic distillation do not introduce strong biases and exhibit superior generalization performance.

\begin{table}[ht]
\caption{Ablation results of distillation loss using VPC metrics in multi-lingual scenarios.}\label{tab:ablation}
\renewcommand\arraystretch{1.3}

\resizebox{1.0\linewidth}{!}{
\begin{tabular}{lccccc}
\hline
         & $\text{EER}_{\text{ig}}$ & $\text{EER}_{\text{la}}$  & WER   & $\rho^{F0}$   & $G_{\textrm{vd}}$   \\ \hline
$\text{MUSA}_\text{ml}$ & 34.88 & 28.76 & 17.06 & 0.73 & -2.73 \\
\quad w/o $\mathcal{L}_{\textrm{sem}}$  & 21.67 & 20.18 & 17.61 & 0.69 & -3.18 \\
\quad w/o $\mathcal{L}_{\textrm{spk}}$  & 24.13 & 22.93 & 25.47 & 0.51 & -3.73 \\ \hline
\end{tabular}
}
\end{table}

\subsection{Ablation Analysis}
MUSA demonstrated strong performance in English, multi-lingual, and unseen language scenarios. We now analyze the effectiveness of the two specially designed distillation methods for insights into their contributions to the overall performance. Table \ref{tab:ablation} shows that when we remove semantic distillation $\mathcal{L}_{\textrm{sem}}$, all metrics degrade, particularly EER results. The decline in EERs suggests that the concealed global representation extracted from the speaker encoder only contains part of the identity information, while other identity information leaks into the residual bottleneck module. These findings indicate that semantic distillation not only allows the residual quantizer layer to model prosody information but also prevents the leakage of global identity information.

Moreover, when we remove speaker distillation $\mathcal{L}_{\textrm{spk}}$ and assume that the first quantizer layer models the linguistic content while the residual quantizer layer models the speaker identity information, we replace the output of the residual quantizer with empty embedding to achieve anonymization. As shown in Table \ref{tab:ablation}, we note that EER results are higher than when removing semantic distillation, and there is a noticeable degradation in WER results and pitch correlation. A possible reason for the performance degradation is that the output of the residual quantizer contains time-varying information, such as prosody, and information loss occurs when it is directly replaced with zero values in the decoder reconstruction process. This highlights the importance of both distillation methods in achieving superior performance in our anonymization approach.

\begin{figure}[ht]
  \centering
  \includegraphics[width=0.9\linewidth]{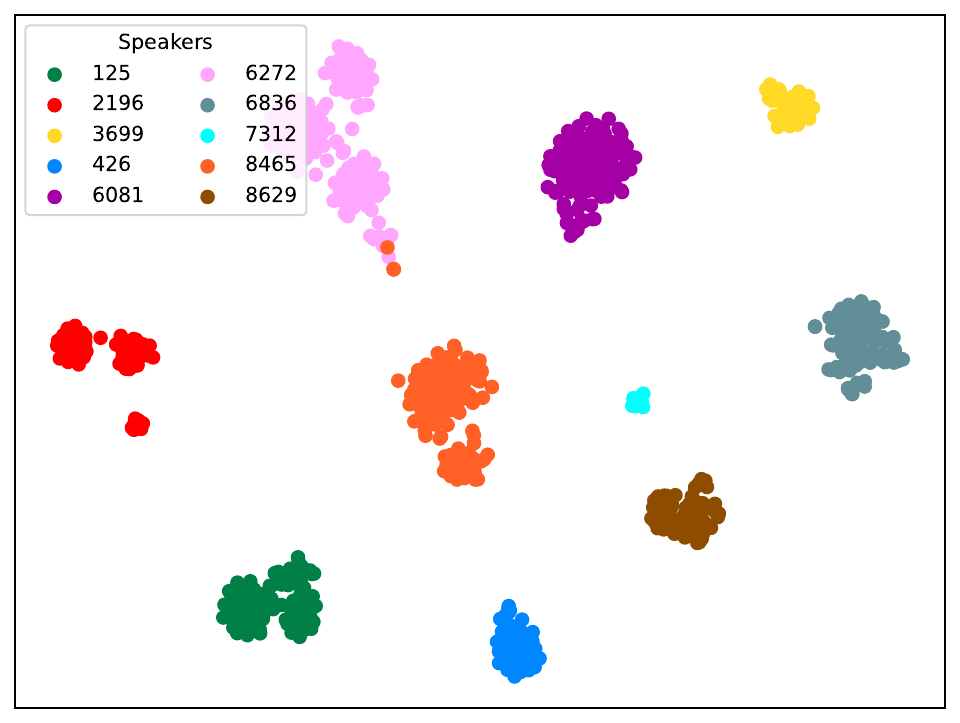}
  \caption{The visualization of the output of the speaker encoder.}
  \label{fig:tsne}

\end{figure}

The speaker anonymization process in our proposed approach is highly dependent on the disentanglement of speaker identity. Therefore, we conduct further visualization of the output of our speaker encoder $E_{s}$. We select 1,147 utterances from 10 random speakers, obtain the global speaker representation using $E_{s}$, and then employ t-SNE for visualization, as shown in Figure~\ref{fig:tsne}. We find that the representation extracted from the speaker encoder can form a distinct cluster for different speakers. This demonstrates that the speaker encoder can capture speaker-related information and further shows the effectiveness of our proposed speaker distillation loss.

\section{Conclusion}
In this study, we present MUSA, a novel anonymization approach capable of anonymizing speech across multiple languages and adapting well to unseen languages. We propose a serial disentanglement strategy using two types of distillation methods to disentangle the global speaker identity representation, linguistic content, and prosody. Based on this serial disentanglement strategy, we propose a simpler and more direct anonymization strategy that conceals the original speaker identity by replacing it with empty embeddings containing zero values. This eliminates the need for conversion to a pseudo-speaker representation and speaker vector pools, thereby reducing complexity. Experiments on VPC datasets and multilingual datasets demonstrate our approach's effectiveness in protecting speaker privacy under various attack scenarios while preserving linguistic content and distinctiveness.

%
\IEEEpeerreviewmaketitle


%

\appendices




\ifCLASSOPTIONcaptionsoff
  \newpage
\fi



\bibliographystyle{IEEEtran}
%
\bibliography{mybib}
\end{document}